\documentstyle[aps,prc,floats,graphicx,tighten]{revtex}

\begin{document}

\twocolumn[\hsize\textwidth\columnwidth\hsize\csname@twocolumnfalse\endcsname

\draft

\title{The preliminary results of fast neutron flux measurements\\
in the  DULB laboratory at Baksan}
\author{J.N.Abdurashitov$^a$, V.N.Gavrin$^a$, A.V.Kalikhov$^a$,
A.A.Klimenko$^{a,b}$, S.B.Osetrov$^{a,b}$,\\
A.A.Shikhin$^a$, A.A.Smolnikov$^{a,b}$,  S.I.Vasiliev$^{a,b}$,
V.E.Yantz$^a$, O.S.Zaborskaya$^a$.}
\address{$^a$Institute for Nuclear Research, 117312 Moscow, Russia}
\address{$^b$Joint Institute for Nuclear Research, 141980 Dubna, Russia}
\date{4 Jan 2000}
\maketitle

\begin{abstract}
One of the main sources of a background
in underground  physics experiments (such as the investigation of solar
neutrino flux, neutrino oscillations, neutrinoless double beta decay,
and the search for annual and daily Cold Dark Matter particle flux
modulation) are
fast neutrons originating from the surrounding rocks.
The measurements of fast neutron flux in the new DULB Laboratory situated
at a
depth of 4900 m w.e. in the Baksan Neutrino Observatory have been performed.
The relative neutron shielding properties of several commonly available
natural
materials were investigated too. The preliminary results obtained with a
high-sensitive fast neutron spectrometer at the level of sensitivity of
about $10^{-7} neutron \ cm^{-2}s^{-1}$ are presented and discussed.
\end{abstract}


] 

\section*{Introduction}

It is well known that one of the main sources of a background
in underground  physics experiments (such as the investigation of solar
neutrino flux, neutrino oscillations, neutrinoless double beta decay, and the
search for annual and daily Cold Dark Matter particle flux modulation) are
fast neutrons originating from the surrounding rocks. The sources of the fast
neutrons are ($\alpha,n$) reactions on the light elements contained in
the rock
(C, O, F, Na, Mg, Al, Si). Neutrons from spontaneous fission of $^{238}$U
take an additional contribution in a total fast neutron flux of about 15-20\%.
Several research groups have investigated the neutron  background at
different underground laboratories \cite{chaz,Arne,alex}. Some of them
used $^6Li$-dopped liquid scintillator technique \cite{alex}, and others
used in addition a Pulse Shape Discrimination technique \cite{chaz}.

The measurements of fast neutron flux in the Deep Underground Low Background
Laboratory of Baksan Neutrino Observatory (DULB BNO) have been performed
with using of a special, high-sensitive fast neutron spectrometer \cite{abdu}.
This laboratory is located under Mt. Andyrchy (Northen Caucasus Mountains,
Russia) in a tunnel that penetrates 4.5 km into the mountain, at a depth of
4900 meters of water equivalent.

The results of such  measurements lead to a conclusion that a neutron
background places a severe limitation on the sensitivity of current and
planned experiments. Owing this fact, the development of new cost-effective,
high-strength radiation shielding against neutrons becomes a very important
task for modern non-accelerator physics experiments. For such purposes
the relative
neutron shielding properties of several commonly available natural materials
were investigated too. Specially, these materials are planned for use in the
construction  of large-volume underground facilities which will be covered
with suitable shielding materials and are situated in the DULB Laboratory at
Baksan.

\section*{Neutron detector}

The spectrometer was constructed to measure low background neutron fluxes
at the level up to $10^{-7} cm^{-2}s^{-1}$ in the presence of intensive
gamma-ray background.

\begin{figure}[t]
\begin{center}
\includegraphics[width=3.375in]{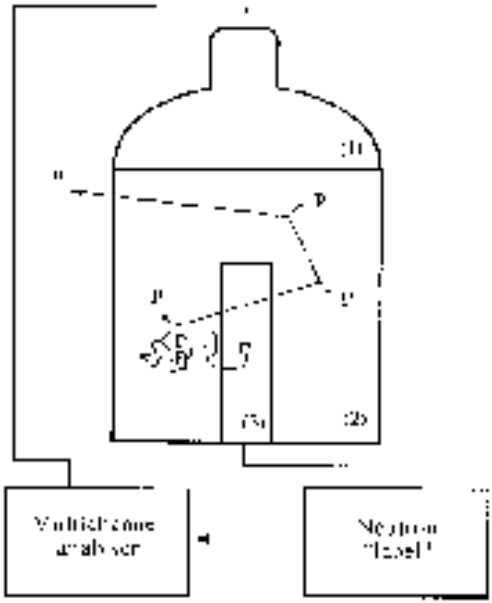}
\end{center}
\caption{The neutron spectrometer schematic view and a principle of
operation. (1) is a PMT, (2) is a liquid scintillator
and (3) is a $^3He$ counter.}
\label{principle}
\end{figure}

The detector consists of 30 l liquid organic scintillator viewed by
photomultipliers with 19 neutron counters ($^3He$ proportional counters)
uniformly distributed through the scintillator volume (see \cite{abdu} for
detail). The spectrometer
schematic view and the principle of operation are shown
in Fig. \ref{principle}.

Fast neutrons with $E_n >$ 1\ MeV entering the liquid scintillator (LS) are
moderated down to thermal energy, producing a LS signal. Then they diffuse
through the detector volume to be captured in $^3He$ counters or on protons
in the scintillator. The LS signal starts the recording system. After
triggering the system waits  a signal from any of the helium counters
for a specific time. This time window corresponds to the delay time between
 correlated events in the scintillator and in the helium counters.
This is one of specific features of the detector. The signal from the LS is
'marked' as a coincident with a neutron capture in the $^3He$ counters only
in the case if a single counter is triggered during the waiting period. An
amplitude of the 'marked' LS signal corresponds to an  initial neutron
energy. This method  allows us to suppress the natural ${\gamma}$-ray
background considerably.

\begin{figure}[t]
\begin{center}
\includegraphics[width=3.375in]{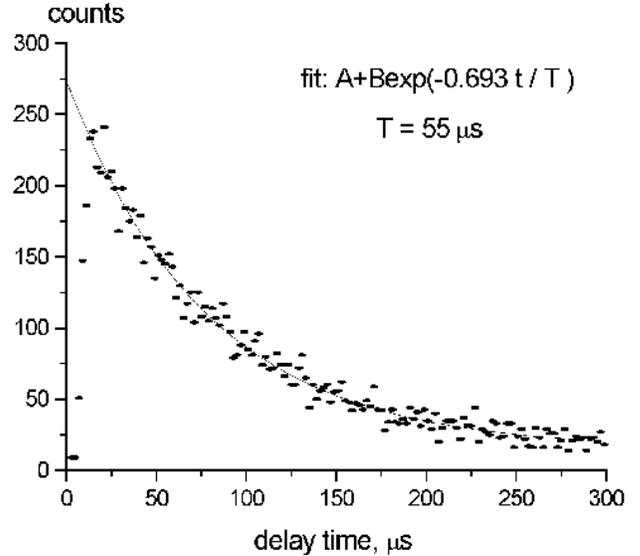}
\end{center}
\caption{Delay time distribution for Pu-Be neutron source}
\label{pbdt}
\end{figure}

The described event discriminating procedure allows us to measure extremely
low neutron fluxes at the level up to $10^{-7} cm^{-2} s^{-1}$ reliably even
if the LS counting rate is as large as several hundred per second. The dead
time of the detector is equal to the delay time (variable value, but
generally
about $120 {\mu}s$) plus about $400 {\mu}s$, which is needed to analyze
a LS event
whether it corresponds to neutron or not. The detection efficiency depends
in a complicated manner on the response function of the detector. As a rough
estimation, we use the value of the efficiency, which is equal to $ 0.04 \pm
0.02$ in the energy range from 1 to 15 MeV. This is based on preliminary
measurements performed with a Pu-Be source. Owing this fact, an absolute
values of the neutron fluxes can be estimated with an uncertainty of  50\%
on the basis of available calibration data. The delay time is a specific
feature
of the detector and depends on the detector design. The acquisition system
allows us to measure the delay time for the neutron events directly. Such
measurements
were carried out using a $Pu-Be$ source with a time window selected to be
equal to $300 {\mu}s$. A typical delay time distribution is shown in Fig.
\ref{pbdt} A fitting procedure leads to a time constant of $T_{1/2} \sim 55
{\mu}s$. According to this result it is sufficient to select the time window
to be equal $120 {\mu}s$ for an actual measurement.

\section*{Measurements}

\subsection{The geometry}

It has been mentioned that we have no yet precise information about the
detection efficiency,
that is why one can calculate absolute value of neutron fluxes with only
$50\%$ certainty. However, it is possible to measure the relative neutron
absorption abilities of various shields with high precision. This information
will be very useful for development of new low background experiments and
searching for cost-effective neutron absoption shields. Such measurements
were carried out in the DULB BNO with using the described neutron
spectrometer.
This new laboratory, consisted of 8 separate counting facilities, is located
under Mt. Andyrchy in a tunnel, which penetrates 4.3 $km$ into the mountain,
at a depth of 4900 $m\ w.e.$

Quartzite and serpentine were selected as materials to be tested because of
comparatively low concentrations of uranium- and thorium-bearing compounds
contained in these rocks. For instant, the measured concentrations of
uranium and thorium for rock serpentine are about
$10^{-8}$g/g in comparison with $10^{-6}$g/g for the surrounding rock. As
for potassium ($^{40}K$) contained in serpentine, it has been found less
than $10^{-8}$g/g in comparison with
$10^{-6}$g/g for the surrounding rock. Measurements of gamma-activity
of different rock samples have been performed with using a well-type
NaI gamma spectrometer with level of sensitivity of about $10^{-9}$g/g,
operated in one of the underground low counting facilities at BNO~\cite{BNO}.
The measured Th, U, and K concentrations in different rock samples
are given in Table \ref{ucon}.

\begin{table}[t]
\squeezetable
\caption{Concentration of U, Th and K in the rock samples.}
\label{ucon}
\begin{tabular}{c c c c}
Sample & $^{238}U, g/g$ & $^{232}Th, g/g$ & $^{40}K, g/g$\\
\hline
Quartzite & $(1.1{\pm}0.1$) $10^{-7}$ & $(4.3{\pm}0.1$)
$10^{-7}$ & ($1.9{\pm}0.03$) $10^{-7}$\\
Serpentine & ($2.2{\pm}0.5$) $10^{-8}$ & $(2.0{\pm}0.9$)
$10^{-8}$  & $<1.2$ $10^{-8}$\\
Surrounding & ($1.6{\pm}0.3$) $10^{-6}$ & $(4.0{\pm}0.1$)
$10^{-6}$ & $(1.6{\pm}0.1$) $10^{-6}$\\
mine rock & & & \\
\end{tabular}
\normalsize
\end{table}

\begin{figure}[t]
\begin{center}
\includegraphics[width=3.375in]{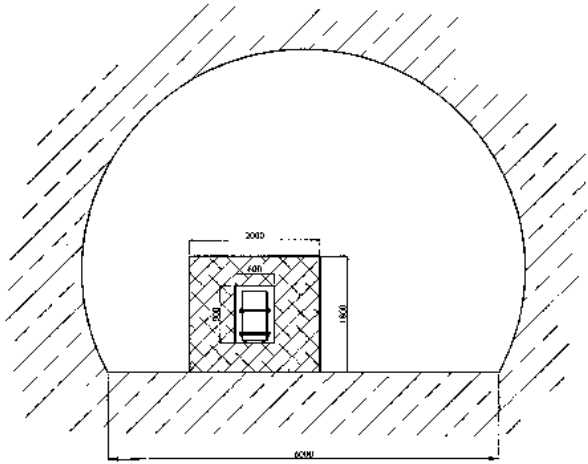}
\end{center}
\caption{Schematic view of cross-section of the DULB experimental site
and the detector inside the neutron shield. All sizes in millimeters.}
\label{lbl}
\end{figure}

Four series of measurements  were performed with the neutron spectrometer
surrounded by different radiation shields. In the first series the
spectrometer was surrounded with a lead shield  4 $cm$ thickness (to reduce
the natural gamma-ray counting rate), and measurements of the natural
neutron background radition field existing in the open experimental site were
performed. In the second and
third series the spectrometer was surrounded with shields of quartzite and
serpentine, respectively.
The rock shields consisted of broken pieces of various sizes, ranging from 1
$cm$ to 15 $cm$, with an effective shield thickness of 35 $cm$ in all
directions. The mean relaxation length of fast neutrons in these shields is
about 15 $cm$ (25 $g/cm^{2}$ for quartzite and 21 $g/ cm^{2}$ for
serpentine). In the fourth series we measured the internal background of the
detector using a neutron-absorbing shield consisted of 40-cm thick section of
polyethylene containing an admixture of boron  and water about 30 $cm$
thick. Schematic
view of one of the investigated neutron shield and cross-section of the DULB
experimental site are shown in Fig.\ref{lbl}.

\subsection{Calibration}

\begin{figure}[t]
\begin{center}
\includegraphics[width=3.375in]{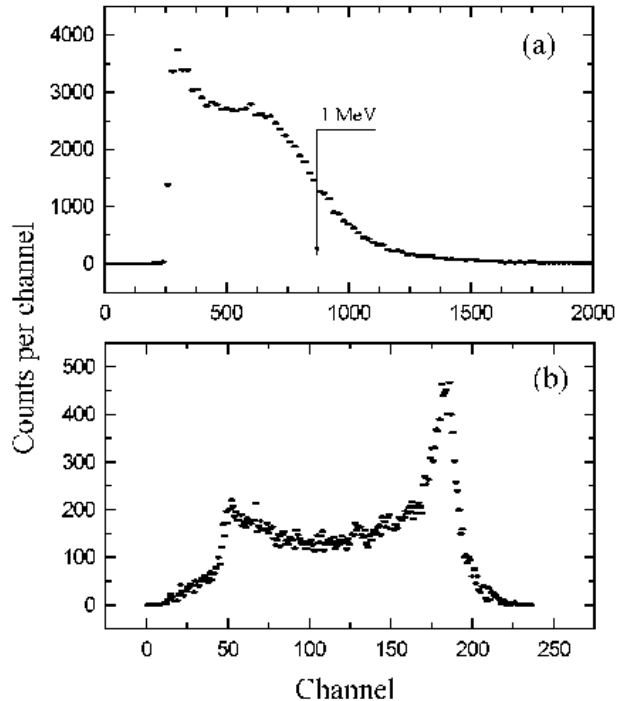}
\end{center}
\caption{Calibration spectra. (a) liquid scintillator irradiated with
$^{60}$Co; (b) $^3$He counters irradiated with Pu-Be source.}
\label{clb}
\end{figure}

A $^{60}Co\ \gamma$-source has been used to calibrate the LS-channel. The
energy of the middle of the Compton edge was assumed to be equal to
1 $MeV$ in the electron energy scale, which corresponds to $\sim 3 MeV$
in the neutron energy scale (see Fig.\ref{clb}a).
A Pu-Be source was used to calibrate the NC-channel of $^3He$ counters.
The spectrum produced by the Pu-Be source in the $^3He$ counters has a
specific shape due to  a wall effect which distorts the counter event spectrum
(see Fig. \ref{clb}b). In spite of this distortion, the range of energies
observed
for true neutron events is less narrow compared to the broad background
spectrum
produced by internal alphas.  Using of the only events from the neutron
window coincident with LS signals makes it possible to suppress the internal
background of the detector.

\subsection{Conditions of measurements}

Main conditions for all series of measurements, such as measuring times, LS-
and NC-counting rates are given in the Table \ref{cond}.

\begin{table*}
\squeezetable
\caption{Conditions of measurements.}
\label{cond}
\begin{tabular}{c c c c c}
Value&No shield&Quartzite&Serpentine&Water+\\
&(5 cm lead)&&&Polyethylene\\
\hline
The measuring time, h&400&290&950&605\\
Dead Time, \%&12&4.3&2.7&1.5\\
Total LS-rate, $s^{-1}$&202&83&62&21\\
Total NC-rate, $h^{-1}$&123 $\pm$ 0.6&103 $\pm$ 0.6&92 $\pm$ 0.3&95$
\pm$ 0.4\\
NC-rate in neutron window, $h^{-1}$&58 $\pm$ 0.4&37 $\pm$ 0.4&27 $
\pm$0.2&27 $\pm$ 0.2\\
Random coincidences rate, $h^{-1}$&1.41{$\pm$}0.005&0.38{$\pm$}0.002&0.19{$\pm$}0.001
&0.07{$\pm$}0.001\\
$R_n$ neutron counting rate, $h^{-1}$  & 29.6 $\pm$ 0.5 &  9.6$\pm$ 0.5 &
-0.2 $\pm$ 0.3 & ---\\
\end{tabular}
\normalsize
\end{table*}

The typical exposure time for each series was a few weeks.  The $\gamma$-ray
background in the open experimental site is high enough that leads to
$\gamma$-counting rate in the LS-channel of about $700\ s^{-1}$. Due to
this fact, following values of dead time were determined for different series:
$12\%$ of the total exposure time for measurements with the lead shield,
$4.3\%$ for quartzite series, $2.7\%$ for serpentine series,
and $1.5\%$ for measurements with the polyethylene/water shield. To calculate
the true neutron counting rates  a proper dead time correction
has been performed.

\section*{Data treatment and results}

Contamination of  $^{222}Rn$ gas inside the experimental site can make a
considerable  contribution \mbox{(up to $20 \%$)} to the  background
$\gamma$-counting rate, which can influence results of the performed
measurements because $^{222}Rn$ activity can vary significantly for a period
of a measurement.

To suppress the count rate variation effect we used a special procedure for
treatment of experimental data. It consists of the following steps.

Two types of data files are stored as a result of a measurement. One of them
contains the information  about neutron energy losses ( LS-signal
amplitudes),
$^3He$ counters signal amplitudes, and delay time for each 'neutron'
candidate
event. Data accumulation was stopped every half-hour and overall numbers
of NC-counts, LS-counts, LS-counts above 1 $MeV$,  and elapsed time
were saved in a file. Total background $\gamma$ - spectra for every
half-hour run were measured
simultaneously and saved in a second file to make it possible to take into
account a time variation of the background $\gamma$-counting rate.

We consider three contributions into the experimentally measured counting
rate $ R_{meas}$: the random coincidence rate $ R_{rnd}$, the internal
detector background counting rate $ R_{bkg}$, and the 'neutron' counting
rate $ R_n$, so that

\begin{equation}
R_n  = R_{meas} - R_{rnd} - R_{bkg}
\end{equation}

\begin{figure}[t]
\begin{center}
\includegraphics[width=3.375in]{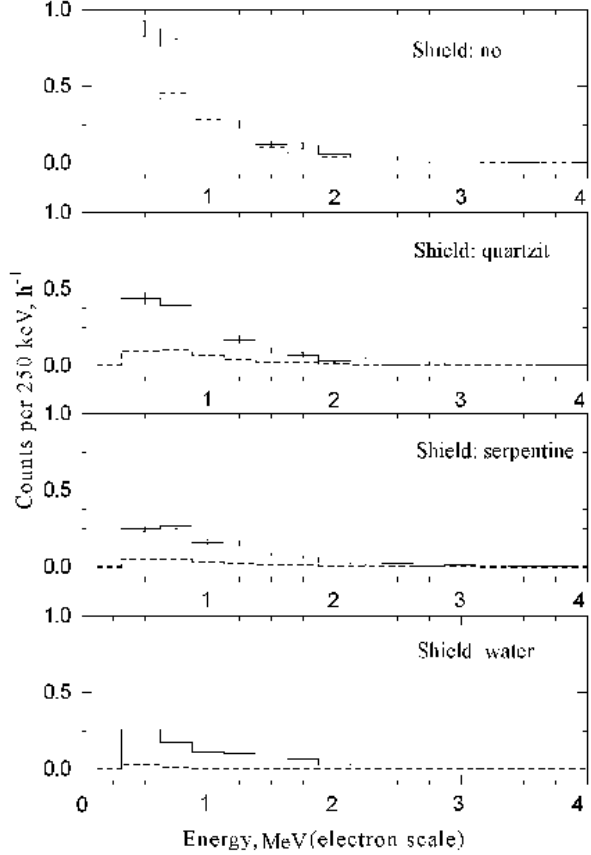}
\end{center}
\caption{The accumulated LS-spectra of all coincidented events ($R_{meas}$,
solid) and the recalculated spectra of random coincidences ($R_{rnd}$,
dashed)}
\label{sprnd}
\end{figure}

We have made obvious assumption that the total background
$\gamma$-spectrum and
the random coincidence spectrum have the same shapes. To obtain random
coincidence spectrum for further subtraction procedure the total background
$\gamma$-spectrum has been normalized with a factor corresponding to the
calculated random coincidence rate. The maximal evaluation for the random
coincidence rate, if  the LS- and NC- events are absolutely independent, can
be calculated by the following way:

\begin{equation}
R_{rand} = r_{\gamma} r^w_n \Delta t,
\end{equation}

\noindent where $r_{\gamma}$ is the $\gamma$-rate, $r_n^w$ is the
$^3He$-counters
counting rate in the determined neutron energy window, ${\Delta}t$ is the
time window. In the case of the performed measurements $(R_{LS}\gg R_{He})$,
this
evaluation is very close to the real counting rate of random coincidences.
Due to a variation in time of  the $^{222}Rn$ activity, the current value of
$r^i_{\gamma}$ depends on time too.
Owing to this fact, we applied the described subtraction procedure to each
half-hour run with corresponding current value of  $ R^i_{rnd}$, and then
summarized resulting neutron spectra in a total serial spectrum.
The accumulated LS-spectra of all coincidented events ($R_{meas}$) and the
recalculated spectra of random coincidence ($R_{rnd}$) for the no-shield,
quartzite, and serpentine series are presented in Fig. \ref{sprnd}.

\begin{figure}[t]
\begin{center}
\includegraphics[width=3.375in]{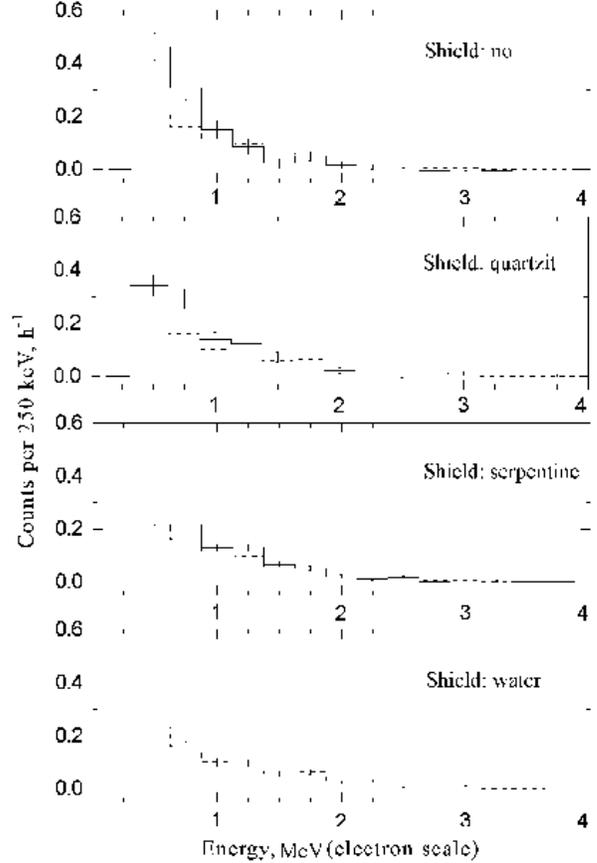}
\end{center}
\caption{Residual LS-spectra ($R_{meas}$ - $R_{rnd}$, solid) in
comparison with the internal background LS-spectrum ($ R_{bkg}$, dashed).}
\label{spbkg}
\end{figure}

An internal detector background spectrum $ R_{bkg}$  has been accumulated
inside the neutron-absorbing
shield consisting of polyethylene and water. Obtained counting rate of the
internal background correlated
(neutron-type, but non-neutron) events was measured as 27 counts per hour,
which in terms of a neutron flux corresponds to ($8.1{\pm}0.5$) $10^{-7}\
s^{-1}cm^{-2}$. The residual LS-spectra ($R_{meas} - R_{rnd}$) in comparison
with the internal background LS-spectrum ($ R_{bkg}$) are presented in
Fig. \ref{spbkg}.

Performing the total subtraction procedure in according with the
equation~(1) we obtain values of the neutron counting rate
$ R_n$ for the no-shield, quartzite, and serpentine series. Taking into
account the detection efficiency uncertainty ($\varepsilon= 0.04 \pm 0.02$)
the obtained values of fast neutron fluxes (above 700\ $keV$ of neutron
energy) are presented here in a following way:

$a (3.5{\pm}1.1)$ $10^{-7}\ s^{-1}cm^{-2}$ for the no-shield measurement,

$a (2.9{\pm}1.1)$ $10^{-7}\ s^{-1}cm^{-2}$ for quartzite shield,

$a (0.6{\pm}0.7)$ $10^{-7}\ s^{-1}cm^{-2}$ for serpentine shield,

were a = ($\varepsilon$ + $\Delta \varepsilon$)/$\varepsilon$.

One can see that the resulting neutron flux measured when the
serpentine shield was in place were found to be at about the minimum level
of sensitivity of the spectrometer. It means that a neutron background
inside the serpentine shield is consisted with
a neuron flux less than $0.7$ $10^{-7} s^{-1}cm^{-2}$. It indicates that
serpentine is indeed clear from uranium and thorium, and is, therefore, the
most likely candidate for use as a cost-effective neutron shield component
material for large-scale low background experiments.

A delay time distribution analysis was performed to understand the
origin of a high level of the internal  detector background.

\section*{Delay time distributions}

\begin{figure}[t]
\begin{center}
\includegraphics[width=3.375in]{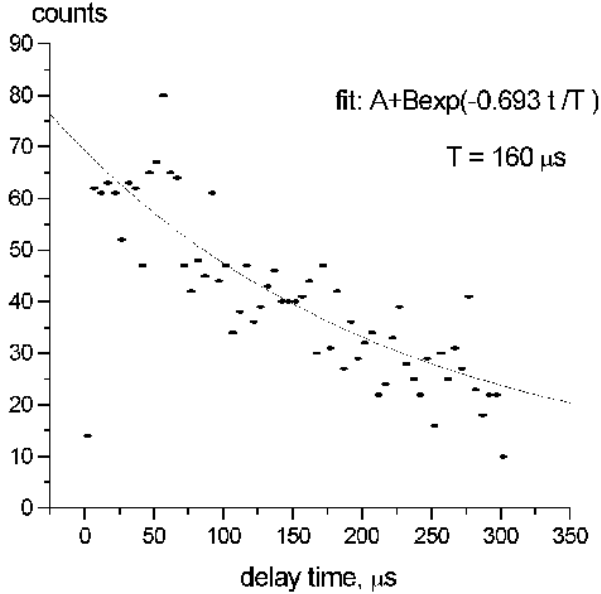}
\end{center}
\caption{Delay time distribution for the coincident events
measured in the water shield.}
\label{wdt}
\end{figure}

Decays of Bi and Po radioactive isotopes, such as
\begin{equation}
^{214}Bi(e,\tilde \nu) \quad \frac{164 \mu s}{ }{\to} \quad
^{214}Po(\alpha) \to {...},
\end{equation}

\noindent which can take place in the helium counter walls, have been
considered as main possible sources of the significant internal background.
To imitate an actual neutron event
beta decay of $^{214}Bi$ can fire the liquid scintillator, followed by a
delayed capture $\alpha$ - signal from Po decay in helium counters.
The delay time distribution of the neutron-type coincidented events
obtained for the series in the
water shield is shown in Fig. \ref{wdt}. Fitting procedure leads to the time
constant $T_{1/2} = 164 {\mu}s$.

\begin{figure}[t]
\begin{center}
\includegraphics[angle=-90, width=3.375in]{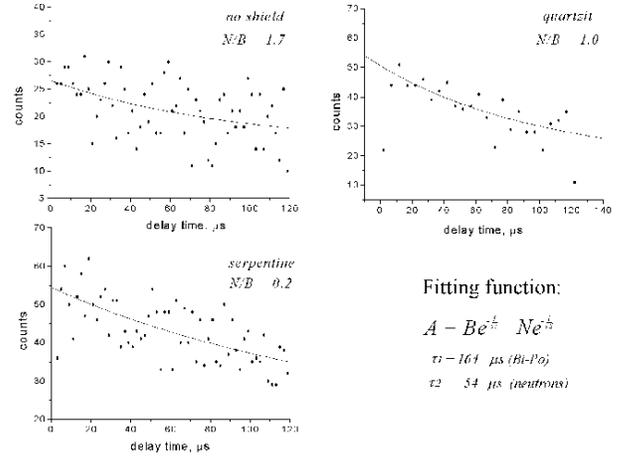}
\end{center}
\caption{Delay time distributions for the coincident events
measured in the series with no shield, quartzite and serpentine.}
\label{dtm}
\end{figure}

It means that, as it was supposed, the origin of the internal background of
our detector is mostly due to contamination of  $^{214}Bi $ in the
$^3He$-counter walls. The delay time distributions for other series
of measurements are shown in Fig. \ref{dtm}.

The following fitting function was used to analize these distributions ($t$
is expressed in ${\mu}$s):

\begin{equation}
A+N e^{-t\ ln2/55}+B e^{-t\ ln2/164},
\end{equation}

\noindent where A is a constant, N is an amplitude corresponding to
neutrons
and B corresponds to internal background. The ratio N/B, which was obtained
in this manner, decreases from measurements in the lead shield
to the measurements in the serpentine shield.

\section*{Conclusions}

The main results of the measurements can be summarized as follows.

(I). The preliminary results obtained from the fast neutron spectrum
accumulated in the open experimental site of the DULB Laboratory at Baksan
is consisted with a neutron flux (for neutrons with energy above 700 keV)
estimated as
values from $5.3\times 10^{-7}\ cm^{-2} s^{-1}$  to $1.8\times 10^{-7}\
cm^{-2} s^{-1}$
depending on the present uncertainty in determination of the detection
efficiency.

(II). The neutron spectrometer sensitivity in a shielded experimental site
is estimated as $0.5\times 10^{-7}\ cm^{-2} s^{-1}$ for a measuring time of
about 1000 h.

(III). It is shown that the main source of the detection sensitivity
limitation, rather then random coincidences, is the internal background of
the spectrometer, which is mostly due to the presence of $\alpha$-particle
emitters ($^{214}Bi - ^{214}Po$ decays) in the $^3He$-counters walls.

(IV). The achieved neutron background inside the serpentine shield is
consisted with a neutron flux less than $0.7$ $10^{-7} s^{-1}cm^{-2}$. It
indicates that serpentine is one of the more likely candidate for use as a
cost-effective neutron shield component material for large-scale
low background experiments.

We have obtained the presented results using the simple event
discrimination procedure and did not use pulse shape discrimination yet.
Nevertheless, it takes us a possibility
to measure extremely low neutron fluxes up to $10^{-7} cm^{-2} s^{-1}$
even when external $\gamma$-counting rate is more than $200\ s^{-1}$.

\acknowledgments

We are grateful to I.I.Pyanzin for the management in proving of reserves and
quarrying of the domestic ultra basic rock samples.
We thank P.S.Wildenhain for careful reading of this article and his critical
remarks. We acknowledge the support of the Russian Foundation of Basic
Research. This research was made possible in part by the grants of
RFBR No.~98--02 16962 and No.~98-02-17973.

\end{document}